# Geometric and electronic properties of two kinds of CrO$_2$ magnetic monolayers: D$_{3d}$ and D$_{2h}$ phases

*Yang Zhang[1], Xianggong Bo[1], Jimeng Jing, Lixia Wang, Shiqian Qiao, Hong Wu, Yong Pu, Feng Li \**

*New Energy Technology Engineering Laboratory of Jiangsu Provence &School of Science, Nanjing University of Posts and Telecommunications (NJUPT), Nanjing 210046, China*

Author Information

*Yang Zhang[1], Xianggong Bo[1]* contributed equally to this work

Corresponding Author

*Feng Li \*:*lifen@njupt.edu.cn

## Abstract

Due to the high magnetic coupling strength between the Cr elements, the bulk phase CrO$_2$ is one of several ferromagnetic oxides known to have the highest Curie temperature. When the dimensionality of the material is reduced from 3D to 2D, the 2D CrO$_2$ system material is expected to maintain a high Curie temperature. In this work, we predict two new phases of CrO$_2$ monolayer ($D_{3d}$ and $D_{2h}$) by using first-principles calculations. We have found that the Curie temperature of 2D CrO$_2$ is much lower than that of its bulk phase, but still remains as high as 191K, which is comparable to that of Fe$_2$Cr$_2$Ge$_6$. In addition, 1L $D_{3d}$-CrO$_2$ is in the ferromagnetic state, while 1L $D_{2h}$-CrO$_2$ is in the antiferromagnetic state. Also, the different geometric structure affects its electrical properties: the 1L $D_{3d}$-CrO$_2$ is a half-metal while 1L $D_{2h}$-CrO$_2$ is a semiconductor. Our studies have shown that there is a wealth of electrical and magnetic properties in CrO$_2$.

## Introduction

Bulk CrO$_2$ is one of several known ferromagnetic oxides, and is the ferromagnetic half-metal that has been experimentally confirmed[1][2][3][4]. It has a rutile structure with Cr ions forming a tetragonal centroid lattice. Cr$^{4+}$ has a closed shell Ar core and two additional 3d electrons. The Cr ions are in the center of the octahedra[5].The saturation magnetization at 10 K was reported to be 1.92 μ$_B$ per Cr site[6], which is close to the ideal value (2 μ$_B$) and is consistent with half-metallicity. It has a high Curie temperature, determined experimentally in the range of 385-400K[7] . In a half-metallic ferromagnet, the conduction electrons should be completely spin polarized. This makes CrO$_2$ a good candidate for use as a spin injector and has sparked the revival of CrO$_2$ thin film growth techniques[7] including the original high-pressure, thermal decomposition method as well as a chemical vapor deposition (CVD) technique discovered in the late 1970s[8].

In the past two decades, with the discovery of graphene[9], the attention of two-dimensional materials has gradually increased. Two dimensional materials have more advantages in spin electronic devices because of their simple structure and easy magnetic control. In 2017, the first intrinsic ferromagnetic two-dimensional CrI$_3$ and CrGeTe$_3$ were successfully fabricated for the first time[10][11], bringing the study of ferromagnetism from 3D down to 2D. When materials are transformed from bulk into their two-dimensional (2D) forms, new physic-al phenomena may appear. Unfortunately, there are few reports on the 2D structure of CrO$_2$ till now. The past studies on half-metal magnets mostly focused on the bulk systems. What necessary to develop low-dimensional half-metal materials is spintronic devices with the demand of small size and high capacity, especially two-dimensional (2D) nanosheets or monolayer materials due to the simple synthesis, ultra-thin thickness, and adjustable electronic structure[12].

In this work, we predicted two lowest-energy structures of CrO$_2$ monolayer, namely 1L $D_{3d}$-CrO$_2$ and $D_{2h}$-CrO$_2$[13][14].They all have high geometric stability, while having significantly different electronic and magnetic properties. The $D_{3d}$-CrO$_2$ has the lowest energy, which is 178.16 meV / cell than that of $D_{2h}$-CrO$_2$. The $D_{3d}$-CrO$_2$ is half-metal ferromagnetic while the $D_{2h}$-CrO$_2$ is an antiferromagnetic semicon-ductor. The both 1L $D_{3d}$-$CrO_2$ and $D_{2h}$-$CrO_2$ could be prepared by advanced chemical vapor deposition method[15][16].

**Methods**

Our first-principles calculations were based on density functional theory (DFT) implemented in the *Vienna Ab initio Simulation Package* (VASP)[17]. Generalized gradient approximation (GGA) for exchange-correlation functional given by *Perdew-Burke-Ernzerhof* (PBE)[18] was used. The effective Hubbard $U_{eff}$ = 2.6 eV was added according to *Dudarev*'s method for the Cr-$d$ orbitals [19][20][21]. The projector augmented wave (PAW)[22] method was used to treat the core electrons. The plane wave cutoff energy was set to be 500 eV. The first Brillouin zone was sampled by using a Γ-centered 20 × 20 × 1 Monkhorst-Pack[23] grid for $D_{3d}$ $CrO_2$ and 10 × 16 × 1 Monkhorst-Pack grid for $D_{2h}$ $CrO_2$. The lattice geometries and atomic positions were fully relaxed until force and the energy were converged to 0.01 eV Å$^{-1}$ and 10$^{-6}$ eV, respectively. Van der Waals interaction (vdW) correlation is considered by using the semi-empirical dispersion-corrected density functional theory (DFT-D3) force-field approach[24]. A vacuum space of 20 Å along the $z$ direction was adopted to model the 2D system. The spin-orbit coupling (SOC) was included in the electronic self-consistent calculations. The phonon dispersion relations was calculated by employing the density-functional perturbation theory (DFPT) and PHONOPY[25]. The Monte Carlo simulation with Wolff algorithhalf-metal based on classical Heisenberg model is used to describe the thermal dynamics of magnetism in equilibrium states[26][27]. All the renorm-ization group Monte Carlo algorithalf-metals were implemented in the open source project MCSOLVER[29].

**Results and discussion**

I $D_{3d}$ $CrO_2$ monolayer

We carefully check energy orders of ferromagnetic states (FM) and antiferromagnetic states (AFM) for the two phases of $CrO_2$. For $D_{3d}$ $CrO_2$, the 4 × 4 × 1 Cr supercells for ferromagnetic state and two antiferromagnetic states are considered to calculate the relative energy (Fig. 1c). The relative energy for these magnetic states are 0, 44.166 and 71.085 meV / unit cell for supercells, respectively. The FM state has the lowest Energy. Therefore, the ground state of $D_{3d}$ $CrO_2$ is ferromagnetic.

The structure of $CrO_2$ monolayer with $D_{3d}$ symmetry is shown in Fig. 1a. For $D_{3d}$ $CrO_2$, the equilibrium lattice constant a and b are 2.92 Å. A Cr atom is surrounded by six O atoms each of them bonded to three Cr atoms. The bond length between Cr and O atom is 1.94 Å. The total magnetic moment is 2 μ$_B$ / unit cell. The majority of the total magnetic moment comes from the Cr atoms, while the O atoms have a small magnetic moment of only 0.18 μ$_B$. The magnetic calculations show that the monolayer behaves as a ferromagnetic half-metal, as shown in Fig. 1b, which is provide a new possibility to implement the intrinsic half-metallicity without any artificial modification. The spin-up channels of $D_{3d}$ $CrO_2$ possess a very large band gap, whereas the spin-down ones do not show any gap. Electrons around the Fermi level are contributed by Cr-$d$ and O-$p$ orbitals. Therefore, the electronic structure has an intrinsic half-metallicity without any artificial modification.

Next, we explore the magnetic behavior under finite temperatures. Here we use the Heisenberg model to describe the magnetic behavior of these systems. The spin Hamiltonian can be written as

$$H = - \sum_{ij} J_{ij} S_i \cdot S_j \qquad (1)$$

where $J$ is the exchange interaction parameter, S = 2/2 for Cr.

Here we considered one typical coupling channel for $D_{3d}$ $CrO_2$ (Fig. 2a). In order to calculate the energy difference between AFM and FM more accurately, we used 2 × 2 × 1 supercells to calculate the exchange coupling strength (Fig. 2b). The energy difference was calculated to be 245.35 meV per 2 × 2 × 1supercell.

According to equation 1, the energies arising from spin exchange for FM and AFM can be written as

$$E_{FM} = E_0 - 12JS^2 \qquad (2)$$
$$E_{AFM} = E_0 + 4JS^2 \qquad (3)$$

The magnetic coupling was calculated to be 15.33 meV. Positive values represent FM exchange interactions. Therefore, the nearest neighbour exchange interaction is ferromagnettic.

Then by classical Metropolis MC simulations(Fig. 2c), the T$_c$ of $D_{3d}$ $CrO_2$ is 191 K.

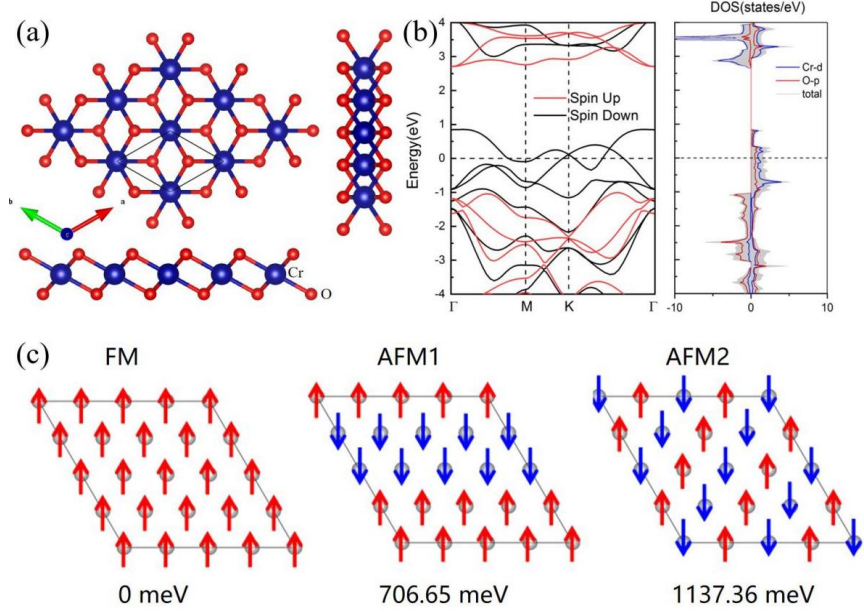

Fig. 1 (a) Structure of $D_{3d}$ CrO$_2$ with top, front and right view. (b) Electronic band structure and density of state of $D_{3d}$ CrO$_2$. (c) The ferromagnetic state and two antiferromagnetic states of $D_{3d}$ CrO$_2$ in $4 \times 4 \times 1$ Cr supercells. Including their relative energy.

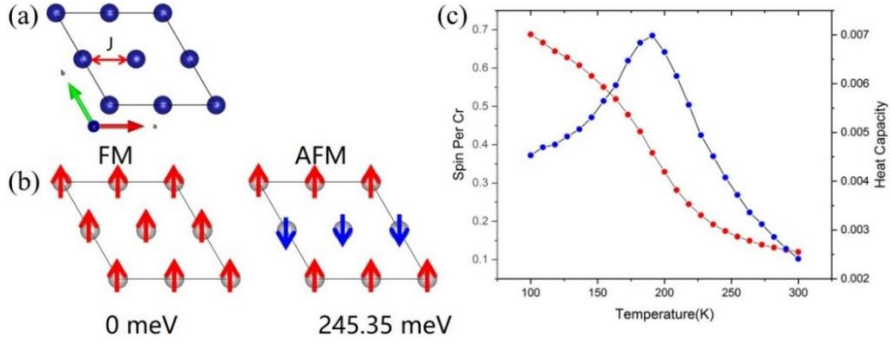

Fig. 2 (a) One classic exchange strength J shown in a $2 \times 2 \times 1$ Cr supercell. For simplicity, O atoms are removed. (b) The ferromagnetic state and antiferromagnetic state of $D_{3d}$ CrO$_2$ in $2 \times 2 \times 1$ Cr supercells. Including their relative energy. (c) MC simulations of T$_c$ of $D_{3d}$ CrO$_2$ monolayer.

## II D$_{2h}$ CrO$_2$ monolayer

Then we discuss the CrO$_2$ monolayer with D$_{2h}$ symmetry. For $D_{2h}$ CrO$_2$, the $2 \times 2 \times 1$ Cr supercells for ferromagnetic state and three antiferromagnetic states are considered to calculate the relative energy (Fig. 3c). The relative energy for these magnetic states are 0, 557.82, 726.37 and 870.32 meV for supercells, respectively. The AFM1 state has the lowest Energy. Therefore, the ground state of $D_{2h}$ CrO$_2$ is antiferromagnetic.

The structure of this phase is shown in Fig. 3a. For $D_{2h}$ CrO$_2$, the equilibrium lattice constant a and b are 5.01 and 2.90 Å, respectively. In a CrO$_2$ unit, there are two typical chemical bond lengths. Four bonds are 1.95 Å and the rest two are 1.93 Å. The magnetic calculations show that the monolayer behaves as an antiferroma-gnetic semiconductor, as shown in Fig. 3b. $D_{2h}$ CrO$_2$ is semiconducting with sizable electronic energy gap (~1.00 eV). The edges of the valence bands are coming from the d orbitals of Cr atoms and p orbitals of O atoms, while the conduction band edges are mainly coming from Cr-$d$ orbitals and partly from the O-$p$ orbitals.

For $D_{2h}$ CrO$_2$, the two Cr atoms have three typical coupling channels whose strengths are labelled as $J_1$, $J_2$ and $J_3$ (Fig. 3a). According to equation 1, the energies arising from spin exchange for FM, AFM1, AFM2 and AFM3 can be written as

$$E_{FM} = E_0 - 8J_1S^2 - 16J_2S^2 - 8J_3S^2 \quad (4)$$
$$E_{AFM1} = E_0 + 8J_1S^2 - 8J_3S^2 \quad (5)$$
$$E_{AFM2} = E_0 - 8J_1S^2 + 8J_3S^2 \quad (6)$$
$$E_{AFM2} = E_0 - 8J_1S^2 + 16J_2S^2 - 8J_3S^2 \quad (7)$$

The magnetic coupling exhibits an anisotropy since the obtained $J_1$, $J_2$ and $J_3$ equal to -42.22, 10.21 and 0.52 meV. The nearest neighbour exchange interaction is antiferromagnetic, the second neighboring and the third neighboring exchange parameters are ferromagnetic.

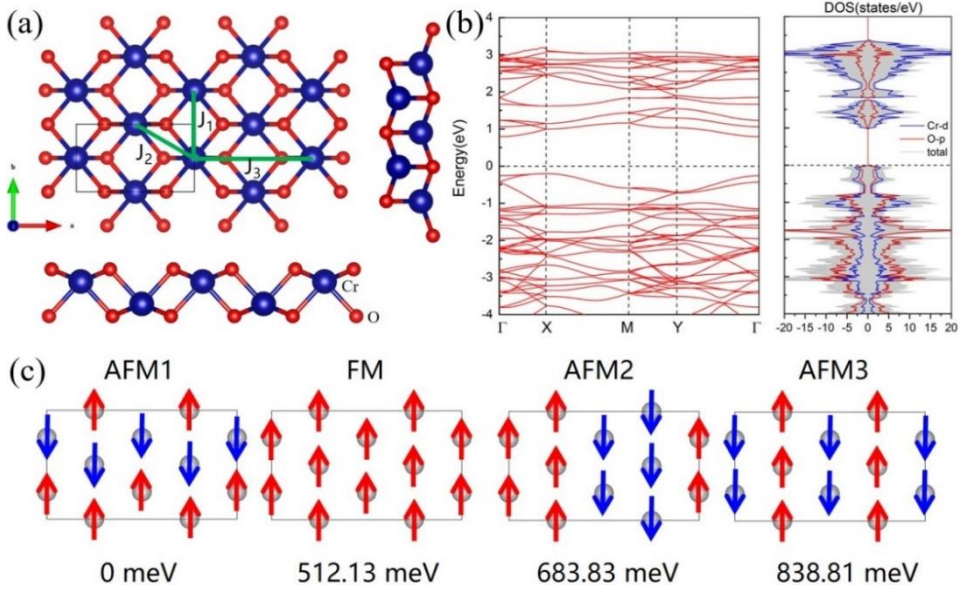

Fig. 3 (a) Structure of $D_{2h}$ CrO$_2$ with top, front and right view. Including three classic exchange strength $J_1$, $J_2$ and $J_3$. (b) Electronic band structure and density of state of $D_{2h}$ CrO$_2$. (c) The ferromagnetic state and two antiferromagnetic states of $D_{2h}$ CrO$_2$ in $2 \times 2 \times 1$ Cr supercells. Including their relative energy.

## III Geometric Stability

To assess the tow phases of CrO$_2$ is dynamically stable, we study its lattice dynamics by calculating the phonon dispersion. The result of $D_{3d}$ is shown in Fig.4a. For monolayer CrO$_2$, the phonon branches is positive in the whole Brillouin zone, so the absence of imaginary mode in the entire Brillouin zone confirms that $D_{3d}$ is dynamically stable. In Fig.4b we can find that one third of the spectral lines around $\Gamma$ are below 0, which is negative. But it is very likely to be stable, maybe because the structural accuracy is not optimized enough, which is also encountered in some other 2D materials, and further improving the optimized accuracy or adding a little tensile strain can eliminate this virtual frequency.

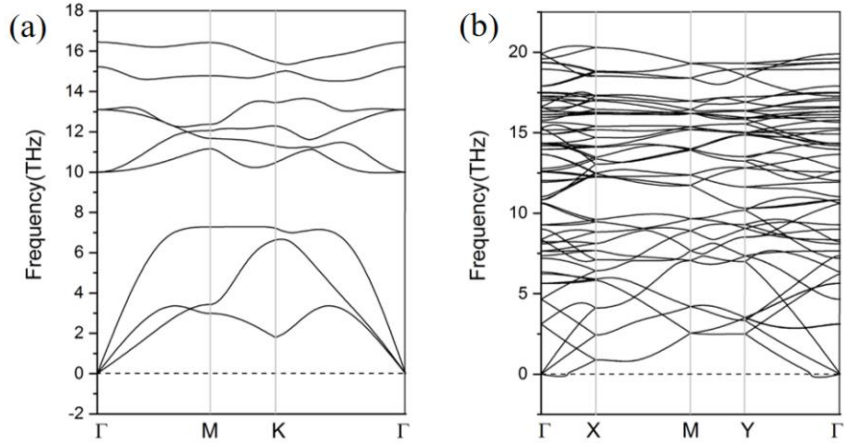

Fig. 4 The phonon spectrum calculations for (a) $D_{3d}$ and (b)$D_{2h}$ CrO$_2$.

## Conclusion

In summary, based on first-principles calculations, we predict two phases of $CrO_2$ monolayer ($D_{3d}$ and $D_{2h}$). Our results indicate that the ground state of $D_{3d}$ $CrO_2$ is FM with estimated $T_c$ of 191 K. And it is an intrinsic half-metal with 100 u % spin polarization, which can be applied in high-quality magnetic recording medium. In contrast, the $D_{2h}$-$CrO_2$ is an antiferromagnetic semiconductor, which renders 2D $CrO_2$ sheet a very promising candidate for AFM spintronics in nanoscale. Our calculations suggest that the two phases of $CrO_2$ can provide a great platform for building spintronic devices and can stimulate relevant experiments.